\documentclass{iau} 
\usepackage{graphicx}

\def \hcm {\hbox {\ifmmode $ atom cm$^{-2}\else atom cm$^{-2}$\fi}}

\def \apj {ApJ}

\def \aap {A\&A}

\def \apss {A\&Sp Sc.}

\def \mnras {MNRAS}

\def \ssr {Space Science Reviews}

\def \aapr {A\&A Rev.}
\newcommand\inte{\textsl{INTEGRAL}}
\title[JD 11.~~HMXBs with INTEGRAL] %% give here short title %%
{Investigating High Mass X-ray Binaries \\ at hard X-rays  with INTEGRAL}

\author[Lara Sidoli \& Adamantia Paizis]   %% give here short author list %%
{Lara Sidoli$^1$
 \and Adamantia Paizis$^1$}

\affiliation{$^1$INAF/IASF Milano, Istituto di Astrofisica e Fisica Cosmica di Milano \\ 
via A. Corti 12, I-20133, Milano, Italy \\ email: {\tt lara.sidoli@inaf.it} and email: {\tt adamantia.paizis@inaf.it} \\[\affilskip]
}

\pubyear{2019}
\volume{346}  %% insert here IAU Symposium No.
\jname{High Mass X-ray Binaries: illuminating the passage from massive binaries to merging compact objects}
\editors{L.~Oskinova, E.~Bozzo \& T.~Bulik, eds.}
\begin{document}

\maketitle

\begin{abstract}
The \inte\ archive developed at INAF-IASF Milano with the available public
observations from late 2002 to 2016 is investigated to extract the X-ray properties of 
58 High Mass X-ray Binaries (HMXBs). This sample consists of sources hosting either 
a Be star (Be/XRBs) or an early-type supergiant companion (SgHMXBs),
including the Supergiant Fast X-ray Transients (SFXTs). \inte\ light curves  
(sampled at 2 ks) are used to build their hard X-ray luminosity distributions, returning the source duty cycles, 
the range of variability of the X-ray luminosity and the time spent in each luminosity state. 
The phenomenology observed with \inte, together with the source variability at soft X-rays taken
from the literature, allows us to obtain a quantitative overview of the main sub-classes of
massive binaries in accretion (Be/XRBs, SgHMXBs and SFXTs). Although some criteria can
be derived to distinguish them, some SgHMXBs exist with intermediate properties, bridging together
persistent SgHMXBs and SFXTs.
\end{abstract}

\firstsection % if your document starts with a section,
\section{Introduction}
High Mass X-ray Binaries (HMXBs) host a compact object (most frequently 
a neutron star [hereafter, NS]) accreting matter from an O or B-type massive star. 
In the great majority of these systems the mass transfer to the 
accretor occurs by means of the stellar wind, 
while in a limited number of HMXBs (SMC~X-1, LMC~X-4, Cen~X-3) 
it happens through Roche Lobe overflow (RLO). 
Before the launch of the \inte\ satellite 
(\cite[Winkler et al. 2003]{Winkler2003}, \cite[Winkler et al. 2011]{Winkler2011}), 
two types of HMXBs were known, depending on the kind of companion, 
either an early type supergiant star (SgHMXBs) or a Be star (Be/XRBs). 

Nowadays the scenario has significantly changed, with a number of Galactic HMXBs tripled 
and new sub-classes of massive binaries  (the ``highly obscured sources''  and 
the ``Supergiant Fast X--ray Transients'', SFXTs) discovered 
thanks to the observations of the Galactic plane performed by the 
\inte\ satellite.
The first type includes HMXBs where the absorbing column density due to the 
local matter is more than one order of magnitude larger than the average 
in HMXBs (reaching 10$^{24}$~cm$^{-2}$ in IGR~J16318--4848).

The SFXTs are HMXBs that 
undergo short (usually less than a few days) outbursts made of 
brief (typical duration of $\sim$2~ks) and bright X--ray flares (peak L$_{X}$$\sim$10$^{36}$~erg~s$^{-1}$),
while most of their time is spent below L$_{X}$$\sim$10$^{34}$~erg~s$^{-1}$. 
The physical mechanism producing this behavior is debated:
the main models
involve different ways to prevent accretion onto the NS
(invoking opposite  assumptions on the NS magnetic field and spin period),
coupled with different assumptions on the donor (clumpy and/or magnetized) wind parameters  
(see \cite[Mart{\'{\i}}nez-N{\'u}{\~n}ez et al. (2017)]{Martinez-Nunez2017}, \cite[Walter et al. (2015)]{Walter2015}
and \cite[Sidoli (2017)]{Sidoli2017review} for  recent reviews, and references therein).

While  SgHMXBs and Be/XRBs differ in the type of their companion star, 
the boundaries between SgHMXBs and SFXTs are 
based {\it only} on their X--ray phenomenology (persistent vs transient X--ray emission),
since they both harbour an early-type supergiant donor. 
In fact, unlike SgHMXBs, SFXTs display a large dynamic range that can reach
six orders of magnitude in X--ray luminosity, between quiescence  and the outburst peak
(as in IGR J17544-2619; \cite[in't Zand 2005]{zand2005}, \cite[Romano et al. 2015]{Romano2015giant}).
For other SFXTs the range of flux variability is typically comprised between 10$^2$ and 10$^4$. 

The present {\em paper}  summarizes our systematic analysis of the \inte\ 
observations of  HMXBs, spanning fourteen years of operations, from 2002 to 2016. 
The main aim of this work is to obtain an overall, quantitative, 
characterization of the different sub-classes of HMXBs 
at hard X--rays (above 18 keV) and to put this phenomenology 
into context of other known properties (like pulsar spin period, orbital geometry) 
and soft X--ray behavior (1--10 keV), as described in the literature. 
We refer the reader to \cite[Sidoli \& Paizis (2018)]{Sidoli2018} for more details on this work.

%%%%%%%%%%%%%%%%%%%%%%%%%%%%%%%%%%%%%%%%%%%%%%%%%%%%%%%%%%%%%%%%%%
\section{The \inte\ archive and the selection of the sample}

Our investigation is based on observations performed by IBIS/ISGRI 
on-board the \inte\ satellite and it is focussed on the
energy range 18--50 keV. 
We built an \inte\ local archive of all public observations 
(see \cite[Paizis et al. 2013]{Paizis2013} and 
\cite[Paizis et al. 2016]{Paizis2016} for the technical details).
For all known HMXBs we extracted the long-term light curves of the sources 
(bin time of $\sim$2~ks, the typical duration of an \inte\ observation, called ``Science Window'') 
spanning fourteen years (from late 2002 to 2016).
We retained in our final sample only the sources which were detected 
(above 5\,sigma) in at least one \inte\ observation 
(i.e. one single Science Window), within 12$^{\circ}$ from the centre of the field-of-view.  
These selection criteria translated into a sensitivity threshold of a few 10$^{-10}$~erg~cm$^{-2}$~s$^{-1}$ 
(18--50 keV) for our  survey, and into a total exposure time of $\sim$200\,Ms for the final HMXB sample.

The final list of sources includes 58 HMXBs, classified in the literature as  
SgHMXBs (18 sources), SFXTs (13 sources) and Be/XRBs (20 sources); the remaining 9 
massive binaries are two pulsars accreting from early type giant stars (LMC~X--4 and Cen~X--3), 
three black hole (candidate) systems (Cyg~X-1, Cyg~X-3, SS~433) 
plus two peculiar massive binaries (IGR J16318-4848 and 3A~2206+543). 
Then, we also included a symbiotic binary XTE~J1743-363, that
is a different kind of wind-fed system, to compare it with massive binaries. 
The complete list of sources is reported in Table 1, together with their sub-class, 
as reported in the literature.

%%%%%%%%%%%%%%%%%%%%%%%%%%%%%%%%%%%%%%%%%%%%%%%%%%%%%%%%%%%%%%%%%%
\begin{table}
  \begin{center}
  \caption{Results of our survey of a sample of HMXBs.}
  \label{tab1}
 {\scriptsize
  \begin{tabular}{|l|c|c|c|c|}\hline 
{\bf Source}    &  {\bf sub-class} & {\bf DC$_{18-50 keV}$ } &   {\bf Av. Luminosity (18-50 keV)$^1$}  & {\bf DR$_{1-10 keV}$}  \\ 
                &                 &   [\%]                  &   [erg s$^{-1}$]                          &  (F$_{max}$/F$_{min}$) \\  \hline
SMC X-1         &    SgHMXB &    49.05            &     1.7E+38           &   7.7 \\
3A 0114+650     &    SgHMXB &    14.63            &     2.1E+36           &   $-$ \\
Vela X-1        &    SgHMXB &    79.22            &     1.3E+36           &   1.7 \\
1E 1145.1-6141  &    SgHMXB &    31.95            &     3.0E+36           &  $-$  \\
GX 301-2        &    SgHMXB &    94.47            &     2.8E+36           &   2.6 \\
H 1538-522      &    SgHMXB &    30.15            &     9.2E+35           &  32.9 \\
IGR J16207-5129 &    SgHMXB &     0.39            &     1.1E+36           &   9.3 \\
IGR J16320-4751 &    SgHMXB &    21.32            &     5.9E+35           &  14.7 \\
IGR J16393-4643 &    SgHMXB &     0.40            &     3.4E+36           &   3.2 \\
OAO 1657-415    &    SgHMXB &    59.78            &     5.8E+36           &  10.0 \\
4U 1700-377     &    SgHMXB &    73.09            &     1.1E+36           &  12.0 \\
IGR J17252-3616 &    SgHMXB &     4.65            &     2.9E+36           &  17.3 \\
IGR J18027-2016 &    SgHMXB &     0.54            &     5.2E+36           &  375  \\
IGR J18214-1318 &    SgHMXB &     0.06            &     3.4E+36           &  $-$  \\
XTE J1855-026   &    SgHMXB &     9.64            &     4.2E+36           &  $-$  \\
H 1907+097      &    SgHMXB &    20.13            &     8.1E+35           &  546  \\
4U 1909+07      &    SgHMXB &    24.84            &     7.1E+35           &  11.5 \\
IGR J19140+0951 &    SgHMXB &    14.18            &     5.2E+35           &  769  \\ \hline
%%%-----------
LMC X-4         &  giant HMXB &  47.23            &     1.2E+38   &      3.4       \\
Cen X-3         &  giant HMXB &  62.79            &     4.0E+36   &      5.0    \\  \hline
%%%-----------
IGR J08408-4503 &    SFXT       & 0.09   &     3.0E+35   &    6750            \\
IGR J11215-5952 &    SFXT       & 0.64   &     1.6E+36   &    $>$480          \\
IGR J16328-4726 &    SFXT       & 0.28   &     1.7E+36   &     300            \\
IGR J16418-4532 &    SFXT       & 1.22   &     6.1E+36   &     308            \\
IGR J16465-4507 &    SFXT       & 0.18   &     2.9E+36   &      37.5          \\
IGR J16479-4514 &    SFXT       & 3.33   &     3.6E+35   &     1667           \\
IGR J17354-3255 &    SFXT       & 0.01   &     3.0E+36   &    $>$929          \\
XTE J1739-302  &     SFXT       & 0.89   &     4.8E+35   &    $>$2040         \\
IGR J17544-2619 &    SFXT       & 0.54   &     5.6E+35   &   1.67$\times10^6$ \\
SAX J1818.6-1703 &   SFXT       & 0.81   &     2.9E+35   &     $>$1364   \\
IGR J18410-0535 &    SFXT       & 0.53   &     3.8E+35   &      1.1$\times10^4$ \\
IGR J18450-0435 &    SFXT       & 0.35   &     1.5E+36   &     513    \\
IGR J18483-0311 &    SFXT       & 4.63   &     5.2E+35   &     899 \\  \hline
%%%-----------
H 0115+634          &   Be/XRB       &     9.55   &     1.5E+37   & 1.4$\times10^5$    \\
RX J0146.9+6121     &   Be/XRB       &     0.11   &     1.1E+35   &    $-$             \\
EXO 0331+530        &   Be/XRB       &    25.10   &     2.4E+37   & 1.07$\times10^6$   \\
X Per               &   Be/XRB       &    76.96   &     2.5E+34   &     10             \\
1A 0535+262         &   Be/XRB       &    12.34   &     4.4E+36   & 2.7$\times10^4$    \\
GRO J1008-57        &   Be/XRB       &     8.87   &     2.4E+36   &    181             \\
4U 1036-56          &   Be/XRB       &     0.35   &     7.5E+35   &     60             \\
IGR J11305-6256     &   Be/XRB       &     0.41   &     1.9E+35   &    $-$             \\
IGR J11435-6109     &   Be/XRB       &     2.68   &     1.4E+36   &    $-$             \\
H 1145-619          &   Be/XRB       &     1.07   &     1.2E+35   &    250             \\
XTE J1543-568       &   Be/XRB       &     0.14   &     2.7E+36   &      8             \\
AX J1749.1-2733     &   Be/XRB       &     0.17   &     8.1E+36   &    $-$             \\
GRO J1750-27        &   Be/XRB       &     4.88   &     2.9E+37   &    $>$10           \\
AX J1820.5-1434     &   Be/XRB       &     0.15   &     2.1E+36   &    $-$             \\
Ginga 1843+009      &   Be/XRB       &     3.39   &     5.8E+36   &    5660            \\
XTE J1858+034       &   Be/XRB       &     5.34   &     8.8E+36   &    $-$             \\
4U 1901+03          &   Be/XRB       &    10.44   &     1.2E+37   &    1000            \\
KS 1947+300         &   Be/XRB       &     9.41   &     6.8E+36   &     800            \\
EXO 2030+375        &   Be/XRB       &    28.99   &     7.8E+36   &  $>$2784           \\
SAX J2103.5+4545    &   Be/XRB       &    11.14   &     2.0E+36   &    6364      \\  \hline
%%%-----------
IGR J16318-4848     & other HMXB     &    35.17   &     7.4E+35   &     3.3        \\
3A 2206+543         & other HMXB     &     6.41   &     2.5E+35   &   250          \\
Cyg X-1             & other HMXB     &    99.88   &     2.5E+36   &     3.7        \\
Cyg X-3             & other HMXB     &    93.49   &     1.0E+37   &     4.9        \\
SS 433              & other HMXB     &    14.97   &     8.5E+35   &     5.0        \\ \hline
%%% ----------
XTE J1743-363       &  symbiotic     &     0.13   &     1.1E+36   &     6.2        \\
\hline
  \end{tabular}
  }
 \end{center}
\vspace{1mm}
 \scriptsize{
 {\it Notes:}\\
  $^1$This 18-50 keV luminosity is an average over \inte\ detections only. 
  This implies that, for transients, it is an average luminosity in outburst. 
  See Sidoli \& Paizis (2018) for the source distances adopted here. 
}
\end{table}
%%%%%%%%%%%%%%%%%%%%%%%%%%%%%%%%%%%%%%%%%%%%%%%%%%%%%%%%%%%%%%%%%%%%%%%%%

\section{\inte\ results}

{\underline{\it Duty Cycles (18-50 keV).}}
The long-term light curves for our sample of HMXBs were used to calculate the source duty cycle in the 
energy range 18-50 keV (DC$_{18-50~keV}$), defined as the percentage of detections 
(at $\sim$2~ks time bin) or, in other words, the ratio between the exposure time when the source 
is detected and the total exposure time at the source position. Table~1 (third column) lists the values obtained. 
Even in case of a persistent SgHMXB, the duty cycle can be lower than 100\%, 
because of source variability leading the source flux below the IBIS/ISGRI threshold of detectability
on the adopted time bin.
Eclipses or off-states also reduce the source duty cycle in persistent sources (e.g. in Vela X-1,
 \cite{Kreykenbohm2008, Sidoli2015vela}).
The advantage of the adoption of a long-term archive, analysed here in a systematic way, 
is that we are confident that the source duty cycles are close to the real source activity, above 
the \inte\ sensitivity.
We refer the reader to Sidoli \& Paizis (2018) for a detailed discussion of the possible 
observational biases.

{\underline{\it Cumulative Luminosity Distributions.}}
The hard X-ray light curves were used to extract the Cumulative Luminosity Distributions (CLDs).
We adopted a single  average conversion factor of 4.5$\times10^{-11}$~erg~cm$^{-2}$~count$^{-1}$ 
from IBIS/ISGRI count-rates to X-ray fluxes (18--50 keV) and assumed the source distances 
reported by \cite[Sidoli \& Paizis (2018)]{Sidoli2018}. 

\begin{figure}[]
%\begin{center}
 \includegraphics[width=2.8in]{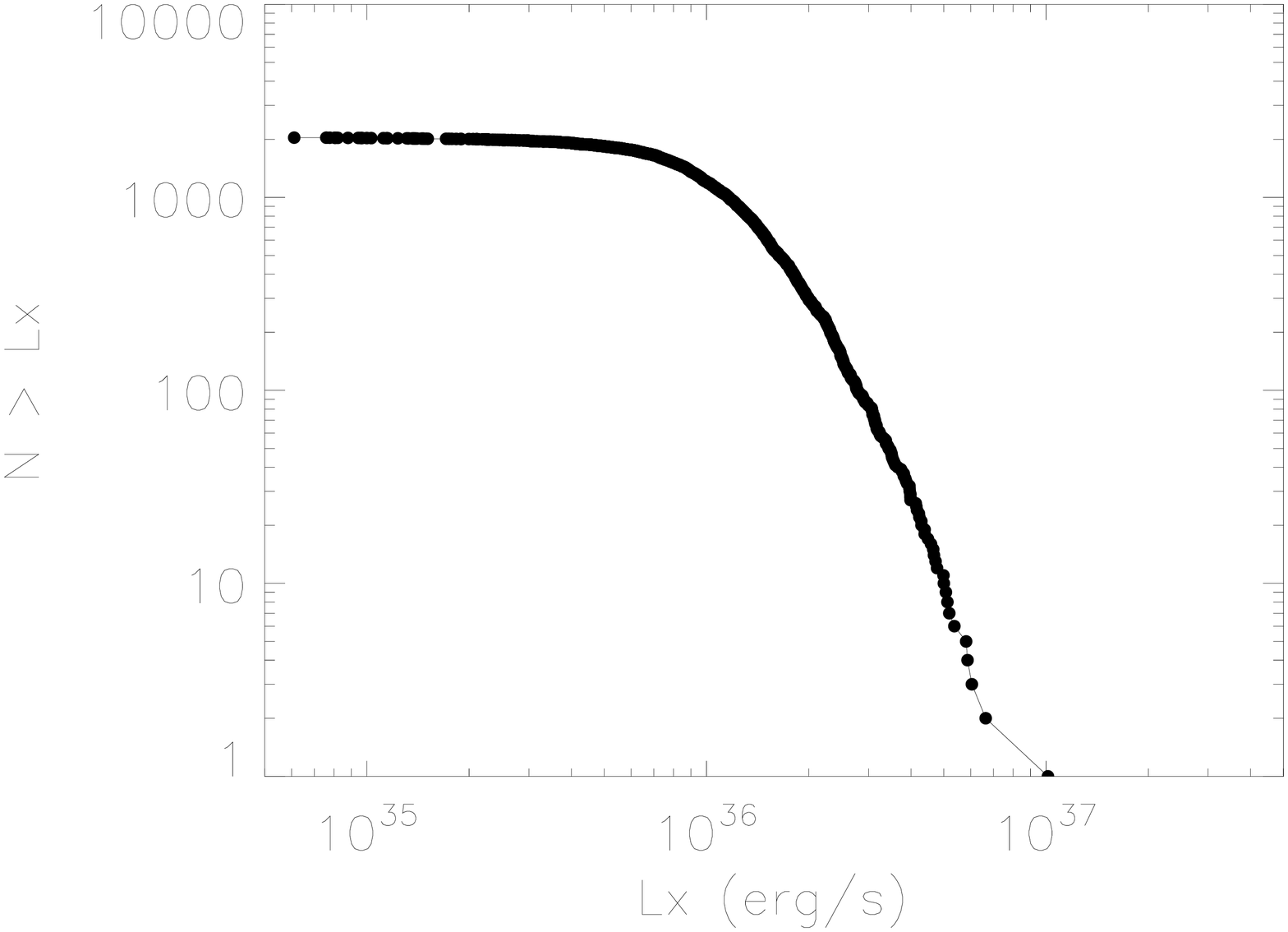} 
 \hspace{-0.9cm}
 \includegraphics[width=2.8in]{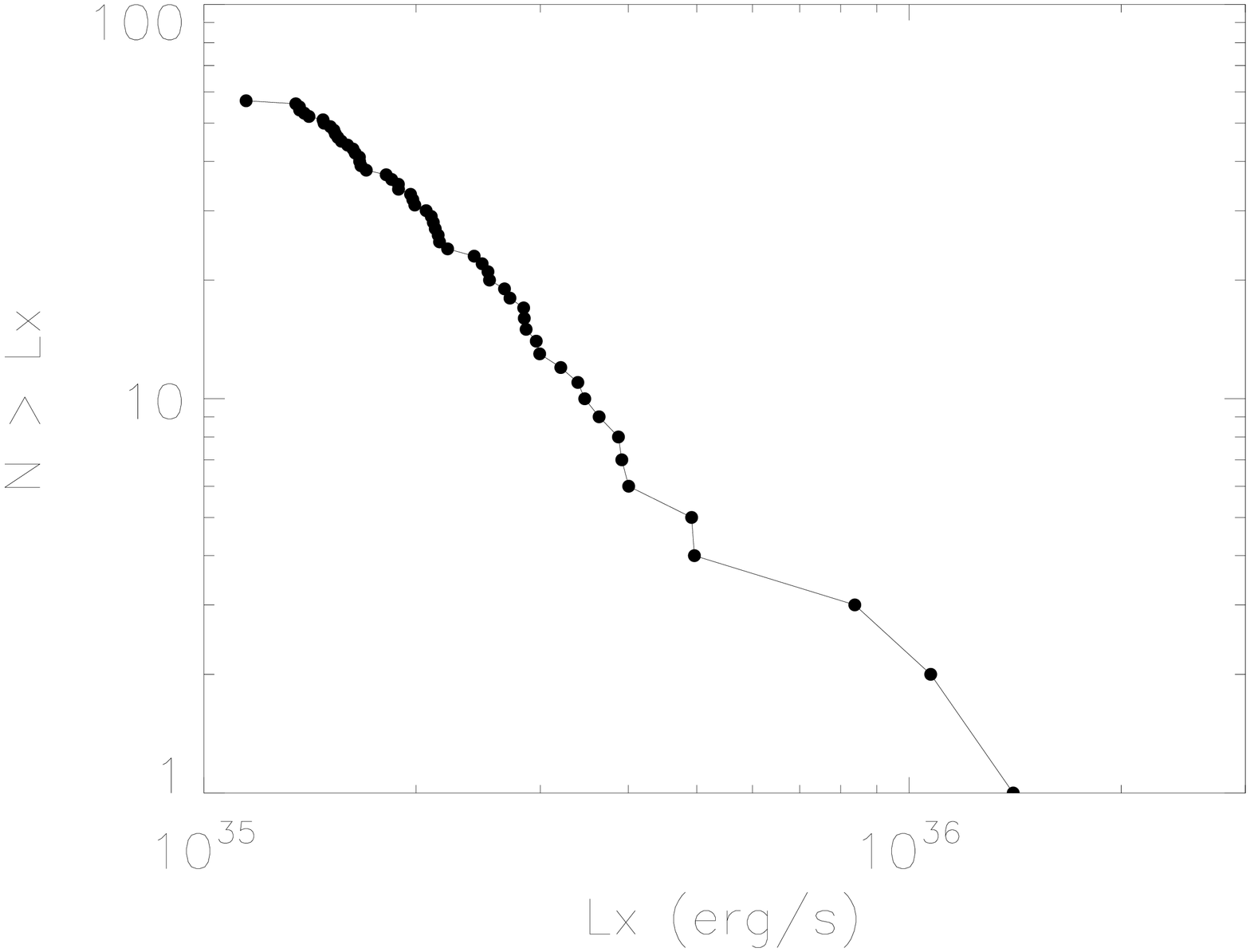} \\
  \includegraphics[width=2.8in]{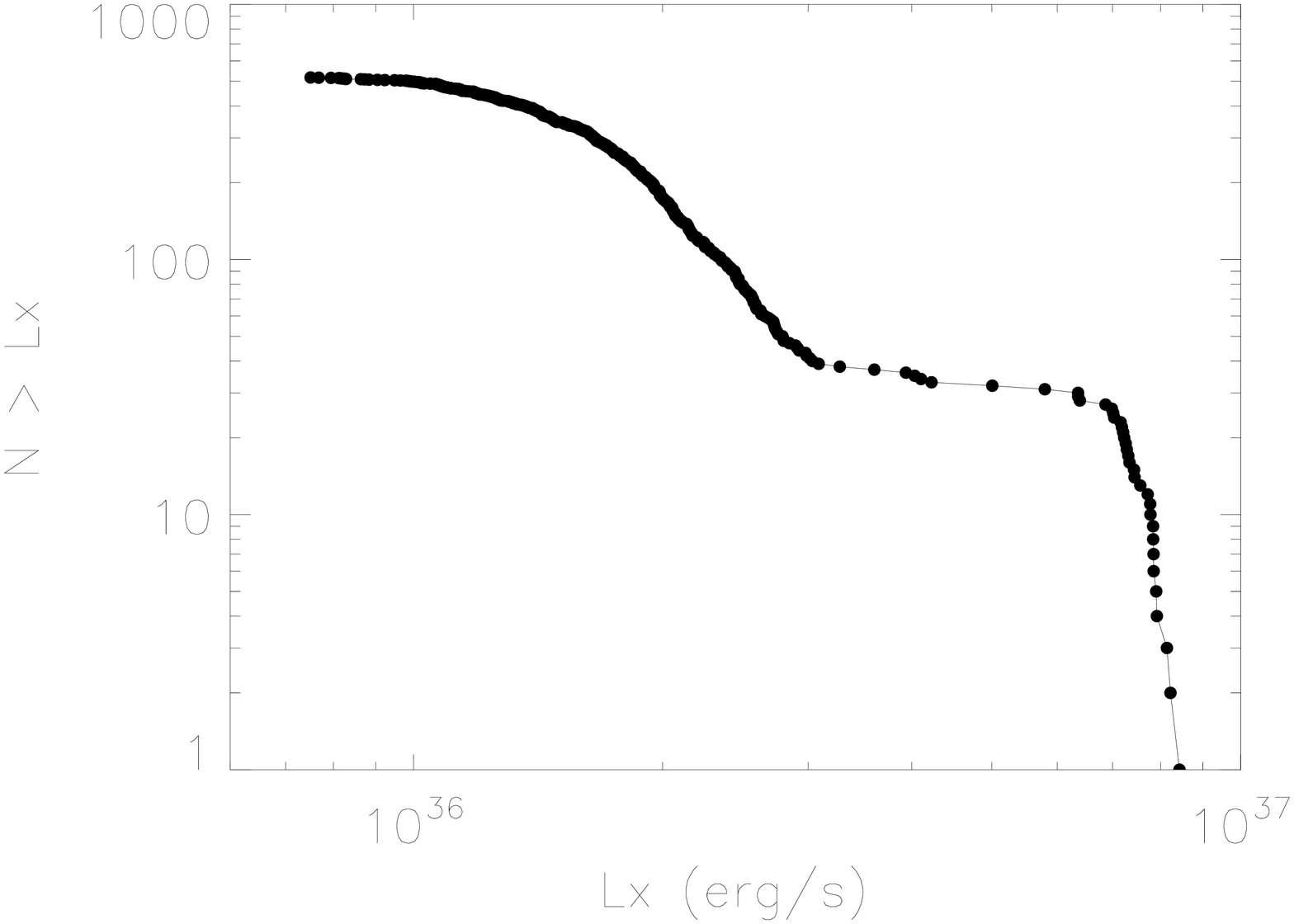} 
  \hspace{-0.9cm}
\includegraphics[width=2.8in]{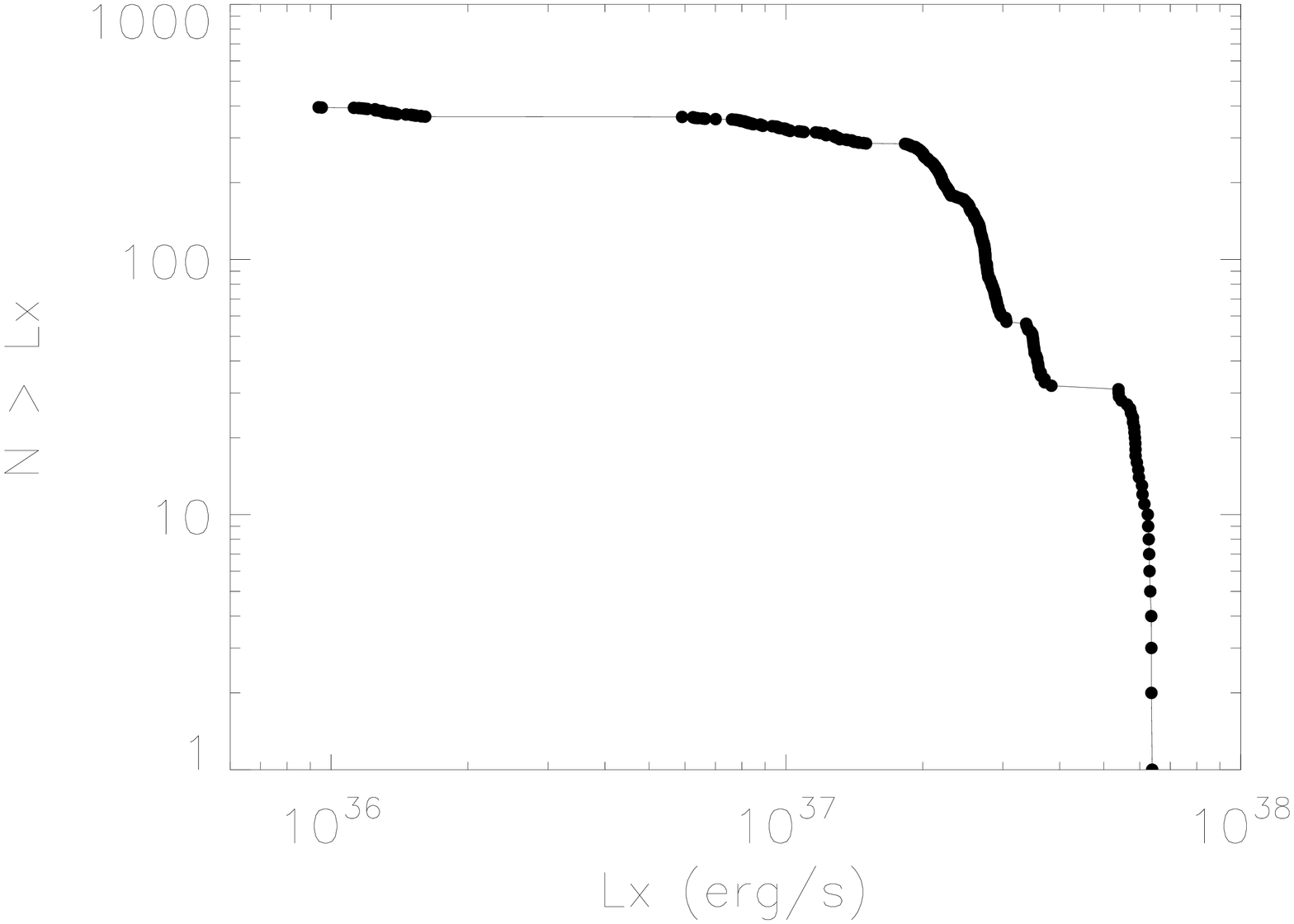} 
 \caption{CLDs of sources representative of the three sub-classes of HMXBs: from left to right, 
 from top to bottom, 
 the persistent SgHMXB Vela~X--1, the SFXT SAX~J1818.6-1703 and the two Be transients SAX~J2103.5+4545 and EXO~0331+530.}
   \label{fig:clds}
%\end{center}
\end{figure}

The CLDs of four sources are shown in Fig.~\ref{fig:clds}, representative of the behavior of 
a persistent SgHMXB (Vela~X-1), a SFXT (SAX~J1818.6-1703) and of two transient Be/XRBs 
(SAX~J2103.5+4545 and EXO~0331+530). Their shape appears different: 
a lognormal-like distribution is evident in Vela X-1, a powerlaw CLD in the SFXT, while 
a more complex behavior is present in the Be/X-ray transients.

Since the timescale of the SFXT flare duration is similar to the bin time of the \inte\ light curves,
the SFXT CLDs are distributions of the SFXT flare luminosities (\cite[Paizis \& Sidoli 2014]{Paizis2014}).
The difference among supergiant systems (SgHMXBs vs SFXTs), between lognormal and powerlaw-like 
luminosity distributions were already found by \cite[Paizis \& Sidoli (2014)]{Paizis2014} from the analysis 
of the first nine years of \inte\ observations of a sample of SFXTs, compared with three SgHMXBs. 

%%%%%%%%%%%%%%%%%%%%%%%%%%%%%%%%%%%%%%%%%%%%%%%%%%%%%%%%%%%%%%%%%%
%%%%%%%%%%%%%%%%%%%%%%%%%%%%%%%%%%%%%%%%%%%%%%%%%%%%%%%%%%%%%%%%%%

%%%%%%%%%%%%%%%%%%%%%%%%%%%%%%%%%%%%%%%%%%%%%%%%%%%%%%%%%%%%%%%%%%
\begin{figure*}[b]
%\vspace*{-2.0 cm}
\begin{center}
%\begin{tabular}{cc}
 \includegraphics[width=4.3in]{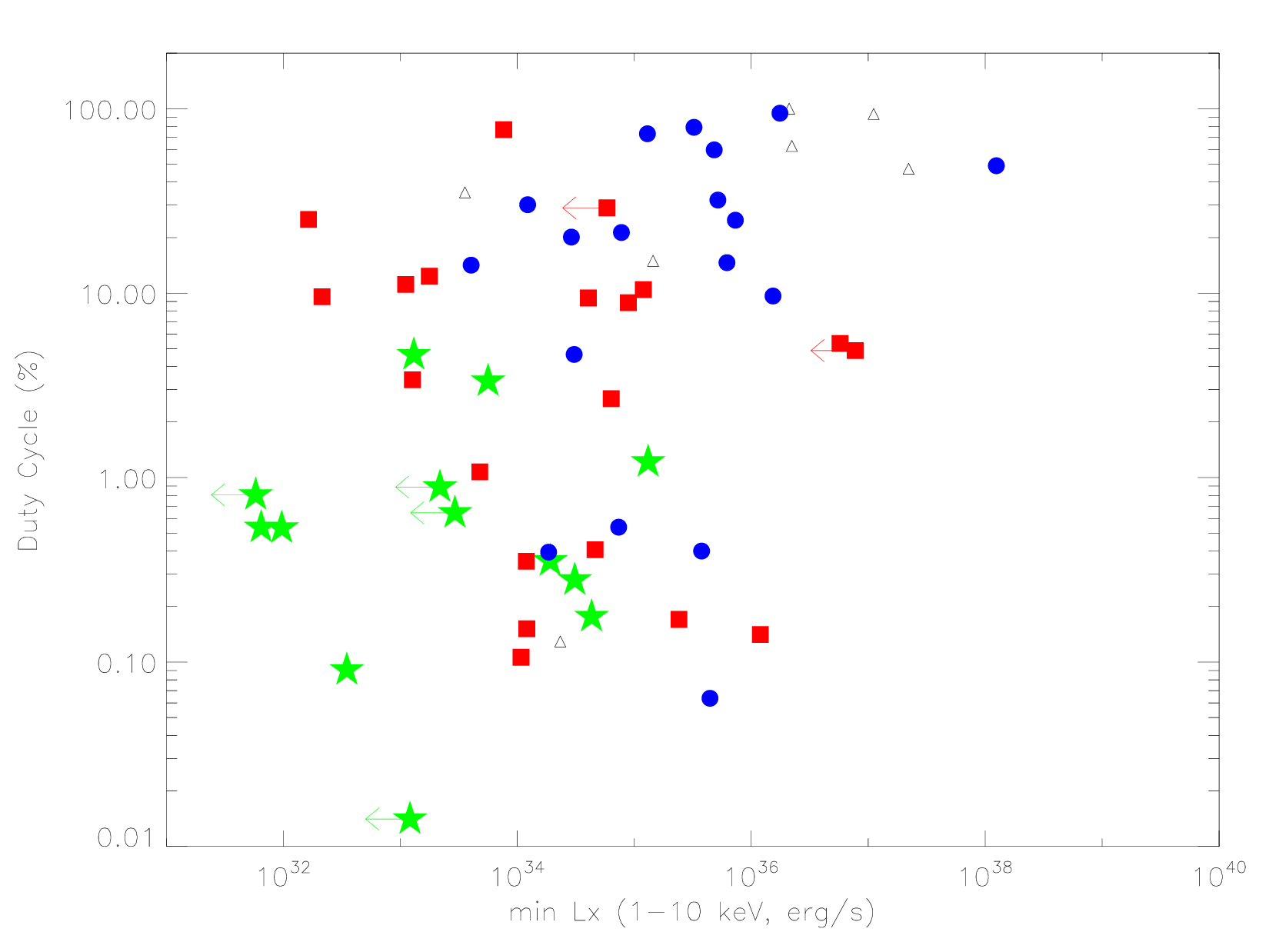} 
 \includegraphics[width=4.3in]{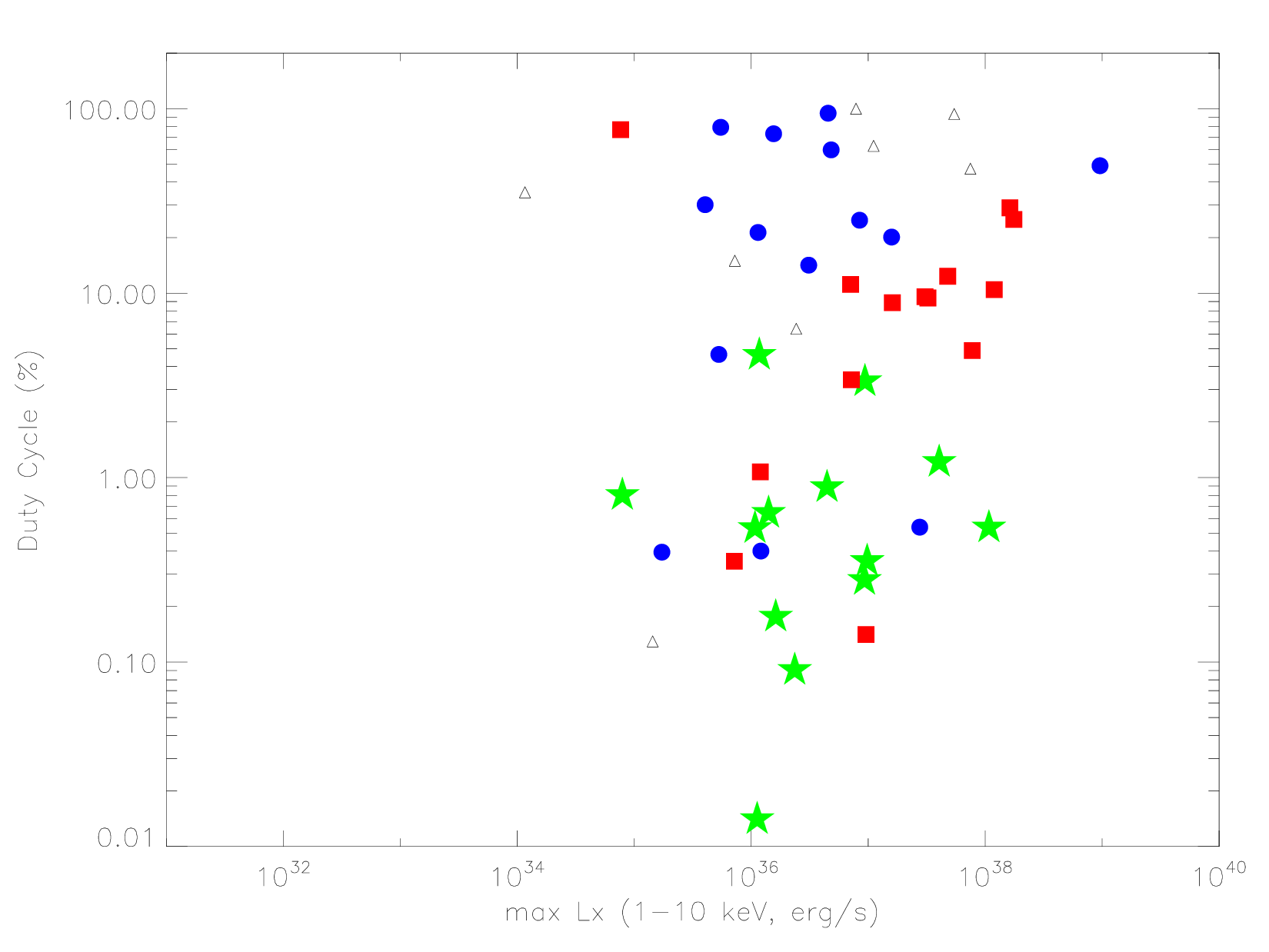} 
% \vspace*{-1.0 cm}
%\end{tabular}
 \caption{\inte\ source duty cycle (DC$_{18-50~keV}$) versus 
 the soft X--ray luminosities reported in the literature 
 (minimum and maximum ones are in the upper and lower panels, respectively). 
 Symbols indicate four sub-classes:  
 blue circles for SgHMXBs, green stars for SFXTs, red squares for Be/XRBs 
 and empty thin diamonds for the remaining types of sources, as reported in Table~1 
 (the ``other'' HMXBs, together with the two giant HMXBs and 
 the symbiotic system). Arrows indicate upper limits on the 
 minimum flux. Note that the different number of sources reported in the 
 two panels are because for some HMXBs we have found in the literature only 
 a single value for the soft X-ray flux  
 (and we ascribed it to the ``minimum 1-10 keV flux''). }
   \label{fig:dr}
\end{center}
\end{figure*}
%%%%%%%%%%%%%%%%%%%%%%%%%%%%%%%%%%%%%%%%%%%%%%%%%%%%%%%%%%%%%%%%%%

%%%%%%%%%%%%%%%%%%%%%%%%%%%%%%%%%%%%%%%%%%%%%%%%%%%%%%%%%%%%%%%%%%
\begin{figure*}[b]
\begin{center}
 \includegraphics[width=4.4in]{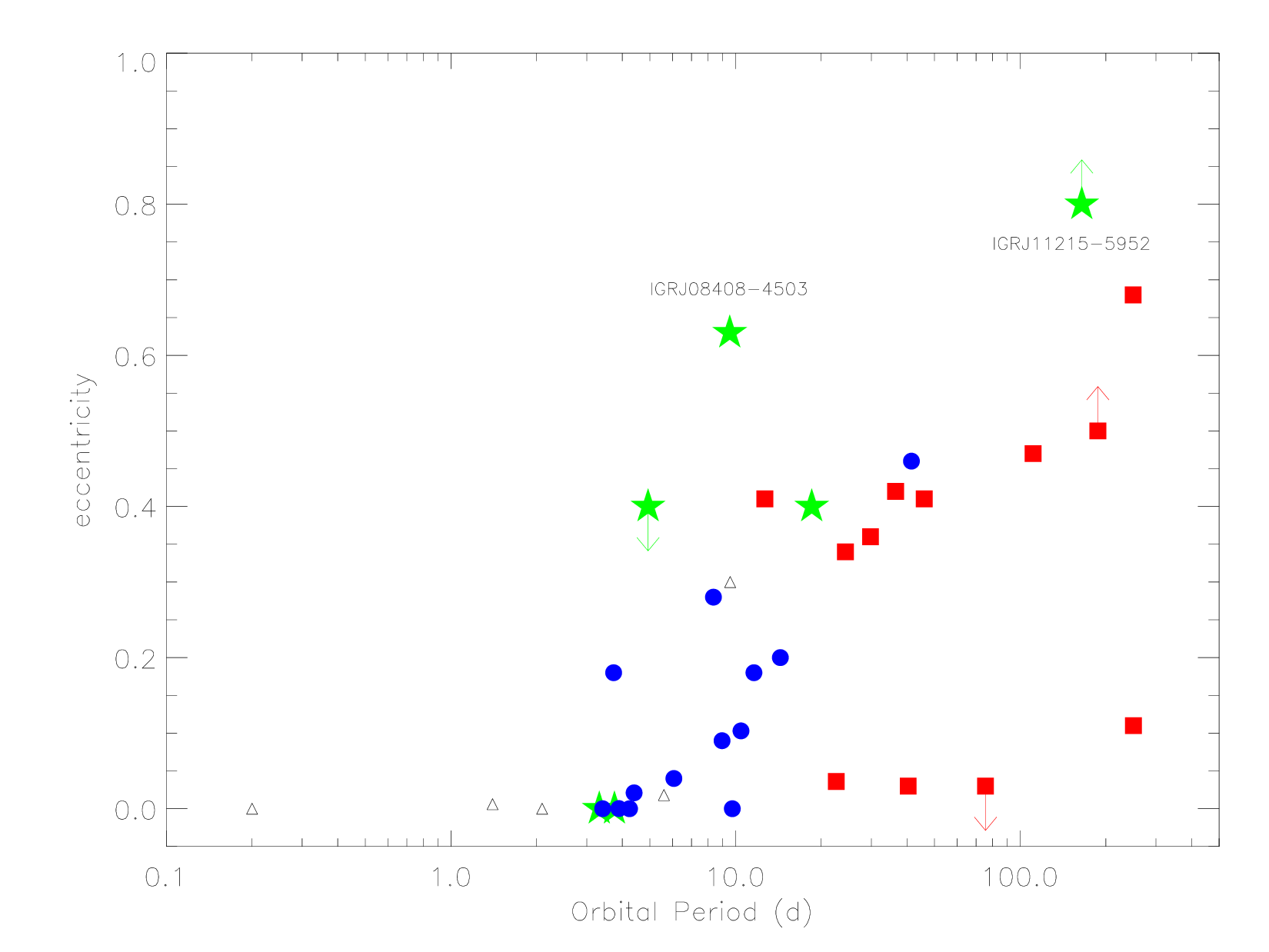} 
 \caption{Orbital eccentricity versus orbital period for our sample of HMXBs. 
 The symbols have the same meaning as in Fig.~\ref{fig:dr}.}
   \label{fig:orbit}
\end{center}
\end{figure*}
%%%%%%%%%%%%%%%%%%%%%%%%%%%%%%%%%%%%%%%%%%%%%%%%%%%%%%%%%%%%%%%%%%
%

This behavior might be ascribed to a separate physical mechanism producing the bright X--ray flares in SFXTs: in the framework of the 
quasi-spherical settling accretion regime (\cite[Shakura et al. 2012]{Shakura2012}), 
hot wind matter, captured within the Bondi radius, accumulates above the  NS magnetosphere;
magnetic reconnection at the base of this shell (between the magnetized, captured, wind matter 
and the NS magnetosphere) has been suggested to 
enhance the plasma entry through the magnetosphere, opening  the NS gate. 
This allows the sudden accretion of the shell material onto the NS and the emission of the SFXT flares
(\cite[Shakura et al. 2014]{Shakura2014}). 
The detection of a $\sim$kG magnetic field from the companion of the SFXT IGR~J11215--5952 
supports this scenario (\cite[Hubrig et al. 2018]{Hubrig2018}).

Transient Be/XRBs can show two types of outbursts, the ``normal'' and the ``giant'' ones
(\cite{Stella1986, Negueruela1998, Negueruela2001, 
Negueruela2001b, Okazaki2001, Reig2011, Kuhnel2015}).
The first type happens periodically and is produced by the higher accretion rate 
when the NS approaches the decretion disc of the Be star, at each passage near periastron.
The second type of outburst can occur at any orbital phase, is more luminous than the normal one and
is thought to be produced by major changes in the Be decretion disc structure.
We ascribe the bimodal behaviour evident in the CLD of SAX~J2103.5+4545 shown in Fig.~\ref{fig:clds} 
to the two different luminosities reached during the two types of outbursts: low (high) luminosity in normal (giant) one, respectively.  
Other Be/XRBs show more complex shapes, multi-modal distributions (like in the case of EXO~0331+530
shown in Fig.~\ref{fig:clds}), 
indicative of multi peaks within the same outburst, or  outbursts reaching different peak luminosities. 
 
The CLDs of all HMXBs of our sample are reported by Sidoli \& Paizis (2018; their Fig.~1-4).
In their normalized version, these functions allow the reader to obtain in one go, 
not only an easy comparison between all kind of HMXBs, but also to quantify the time spent 
by each HMXB in any given luminosity state, above the instrumental sensitivity.

%%%%%%%%%%%%%%%%%%%%%%%%%%%%%%%%%%%%%%%%%%%%%%%%%%%%%%%%%%%%%%
{\underline{\it Average Luminosity (18-50 keV).}}
An average luminosity (18-50 keV) was calculated for each source, over
the \inte\ detections (at 2~ks timescale; see Table~1, forth column).
Note that this definition implies that, for transient sources, this 
is an average luminosity \textit{ in outburst}.

%%%%%%%%%%%%%%%%%%%%%%%%%%%%%%%%%%%%%%%%%%%%%%%%%%%%%%%%%%%%%%%%%%
\begin{figure}[]
\begin{center}
\includegraphics[width=4.0in]{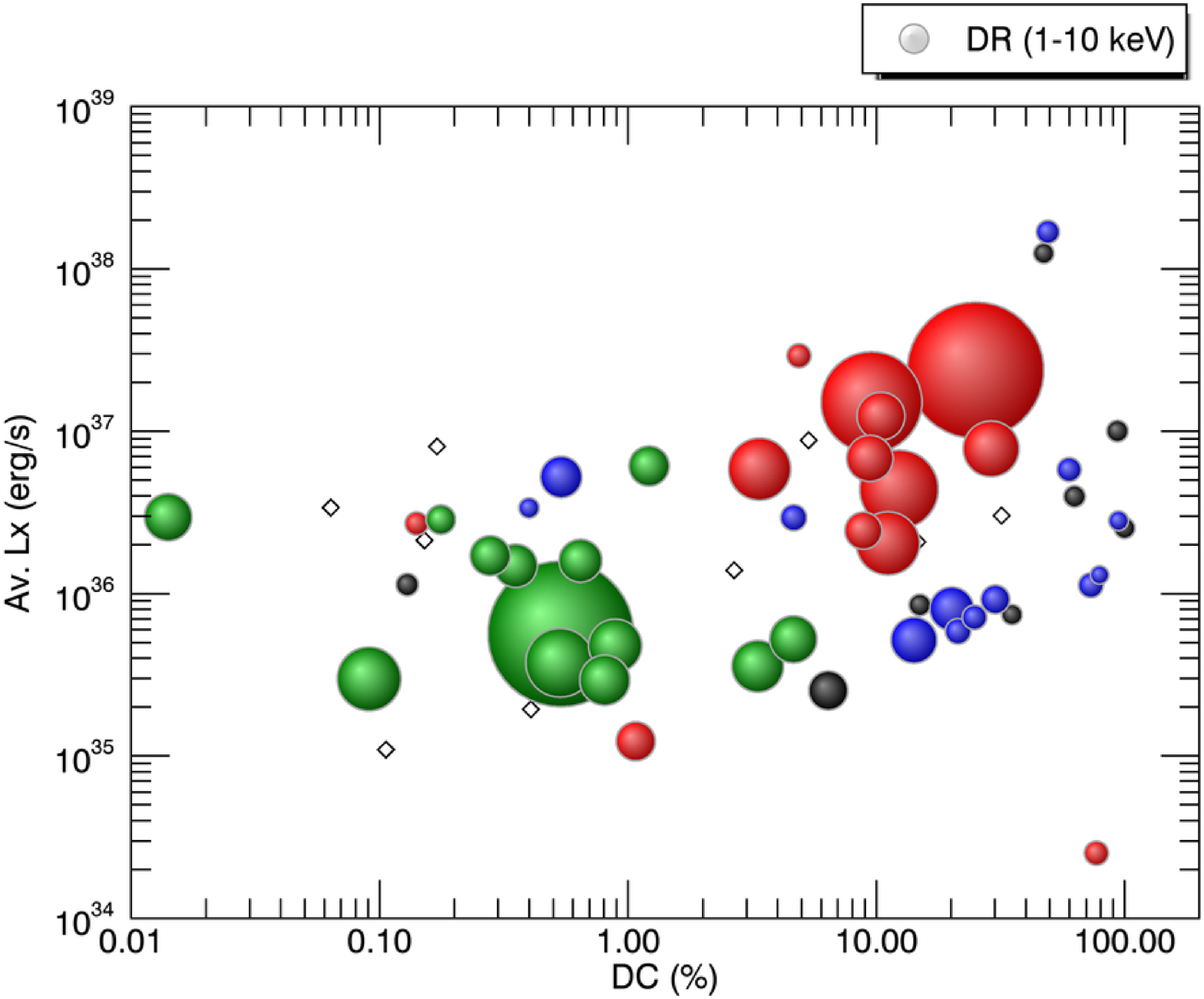} \\
 \includegraphics[width=4.0in]{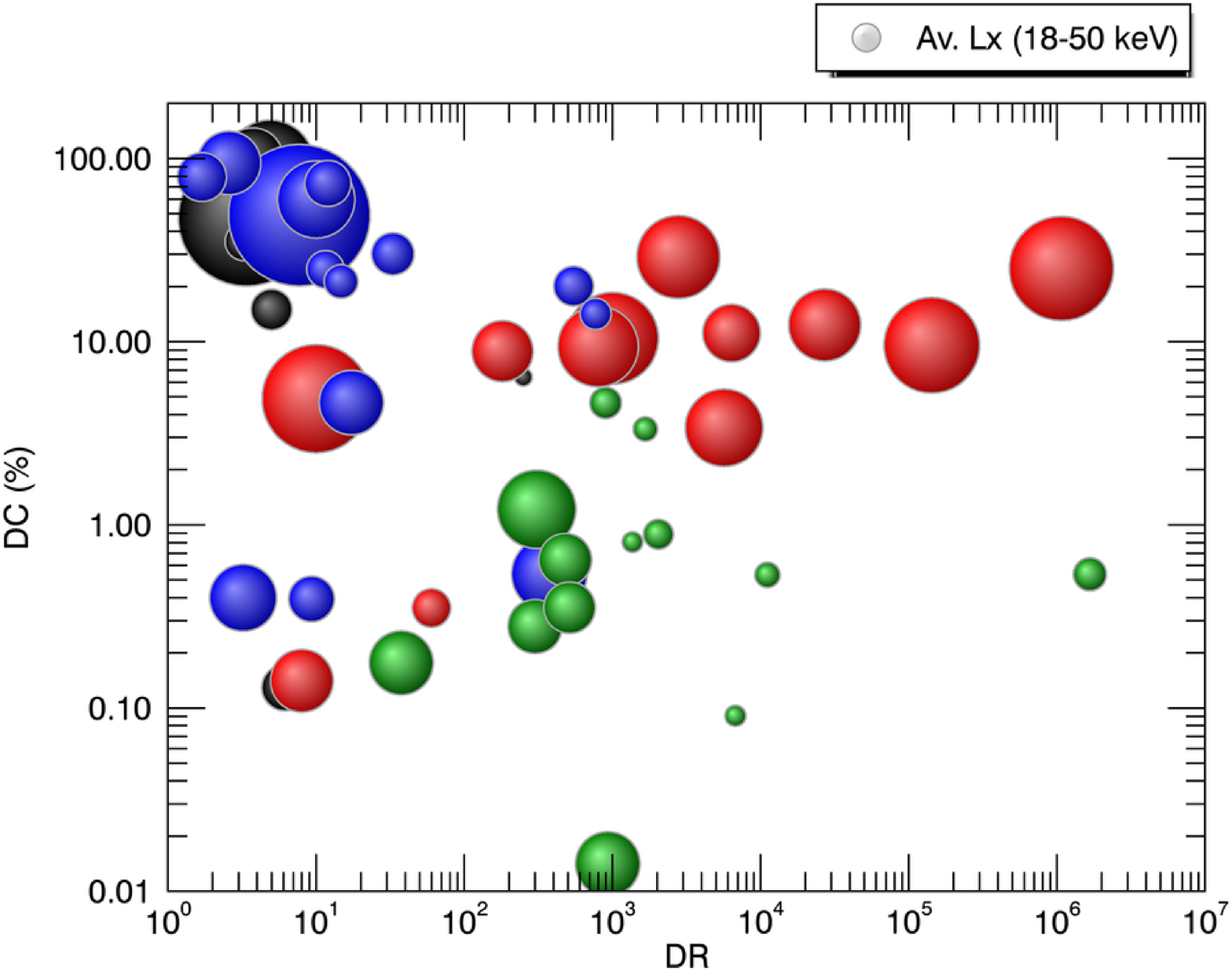} 
 \caption{Bubbleplots summarizing the results obtained, focussing on three main quantities:  
DC$_{18-50\,keV}$, DR$_{1-10\,keV}$ and the average luminosity in the energy range 18--50\,keV 
(note that, for transient sources, this is a luminosity during outburst, as observed by \inte). 
\textit{Upper panel}: average L$_{18-50\,keV}$ vs DC$_{18-50\,keV}$; 
each bubble indicates a single source, with the bubble size correlating 
with the DR$_{1-10\,keV}$. The diamonds mark the position of sources for which 
the DR$_{1-10\,keV}$ could not be calculated (only a single flux in the 1-10\,keV energy band 
was found in the literature).  In blue we mark the SgHMXBs, in red the Be/XRBs and in green the SFXTs. 
In black, all other systems are indicated
(the ``other" HMXBs reported in Table~1, together with the giant HMXBs and the symbiotic binary).  
\textit{Lower panel}: DC$_{18-50\,keV}$ vs  DR$_{1-10\,keV}$, with bubble sizes 
correlating with  the average L$_{18-50\,keV}$.}
   \label{fig:bubble}
\end{center}
\end{figure}
%%%%%%%%%%%%%%%%%%%%%%%%%%%%%%%%%%%%%%%%%%%%%%%%%%%%%%%%%%%%%%%%%%

\section{Other HMXB properties from the literature}

{\underline{\it Dynamic Ranges (1-10 keV).}} 
Other source properties were collected from the literature, in order
to put the \inte\ behavior into a wider context:
distance,  pulsar spin and orbital period, eccentricity of the orbit, maximum and minimum fluxes
in soft X--rays (1--10 keV, corrected for the absorption). 
These latter were investigated since the instruments observing the sky at soft X--rays 
are much more sensitive than \inte\ and can probe the true quiescent state in transient sources, 
together with their variability range between quiescence and outburst peak. 
 
When the published soft X--ray fluxes were not available in the 1--10 keV range,  
we extrapolated them using  \textsc{WebPIMMS} and the appropriate model found in the literature. 
Then, we calculated their ratio (the dynamic range ``DR$_{1-10~keV}$'' = F$_{max}$ /  F$_{min}$, reported 
in Table~1, last column). 
When only a single value for the soft X-ray flux was found, the 
dynamic range was not calculated (``$-$'' in Table~1) 
and the flux was ascribed to the ``minimum flux". 
Note that we considered only spin-phase-averaged fluxes for X-ray pulsars,  
and  out-of-eclipse minimum fluxes for eclipsing systems, to obtain the intrinsic range of 
X--ray variability.  

In Fig.~\ref{fig:dr} we show the source DC$_{18-50~keV}$ plotted against
the minimum and maximum luminosities (1--10 keV), for different HMXB sub-classes: 
the scatter is huge in the upper panel where the duty cycle is plotted versus the minimum
soft X--ray luminosity. The SFXTs are located in the lower left part of the
plot, at low DC$_{18-50~keV}$ and 
X--ray luminosity in quiescence, while the persistent SgHMXB mostly lie 
in the upper right part, at both high luminosities and large duty cycles. 
Be/XRBs appear located in-between them. 
In the lower panel, where the maximum soft X--ray luminosity is considered, 
the sub-classes regroup to the right, at more similar luminosities 
(in outbursts for SFXTs and Be/X-ray transients). 
A few sources, classified in the literature
as SgHMXB (blue circles in Fig.~\ref{fig:dr}),  display a very low DC$_{18-50~keV}$, 
similar to SFXTs. They might be either mis-classified transients or 
persistent sources emitting X--rays at a level just below the 
instrumental sensitivity, that are detected only during sporadic flaring. 
Note that the HMXBs almost reaching the  Eddington luminosity 
are the RLO systems SMC X--1 and LMC~X--4.

{\underline{\it Orbital geometry.}} 
Among the many trends  of source properties we have investigated for our sample 
(see Sidoli \& Paizis, 2018), we report here on the plot showing
the system eccentricity versus the orbital period (Fig.~\ref{fig:orbit}).
Two trends are evident, above P$_{orb}$$\sim$10~d: low eccentricity Be/XRBs with no correlation 
with the orbital period (X~Per is the prototype) and a group of binaries (mostly Be/XRBs) 
where the eccentricity correlates with the orbital period. 
SgHMXBs are located at lower eccentricities and orbital periods. 
This plot has already been investigated in the literature (\cite[Townsend et al. 2011]{Townsend2011}). 
The novelty here is the inclusion of SFXTs (not considered by Townsend et al. 2011): some of them display
circular orbits, while others very eccentric geometries, like
IGR~J08408-4503 (e=0.63 and P$_{orb}$=9.54~d) and 
IGR~J11215-5952 (e$>0.8$ and P$_{orb}$=165~d).
These SFXTs enable the HMXBs hosting supergiant stars 
to extend at larger eccentricities and orbital periods, in a parameter space that is 
unusual even for Be/XRBs.

\section{Conclusions}

We summarize  the results of our systematic analysis in Fig.~\ref{fig:bubble}, 
making use of three characterizing quantities: two of them have been 
derived from the analysis of fourteen years of \inte\ observations  
(DC$_{18-50~keV}$ and the average 18--50 keV luminosity, in outburst for transients), 
while the third one has been calculated from 
soft X--ray fluxes taken (or extrapolated) from the literature (DR$_{1-10~keV}$).

We have obtained a global view of a large number of HMXBs where the different kind of sources
tend to cluster mainly in different region of this 3D space, as follows:

\begin{itemize}
 
\item SgHMXBs (excluding the high luminosity RLO systems) in general show 
low DR$_{1-10~keV}$ ($<$ 40), 
high duty cycles (DC$_{18-50~keV}$$>$10 per cent),
low average 18--50 keV luminosity ($\sim$10$^{36}$~erg~s$^{-1}$);

\item SFXTs are characterized by 
high DR$_{1-10~keV}$ ($>$100),   
low duty cycles (DC$_{18-50~keV}$$<$5 per cent),
low average 18--50 keV luminosity in outburst ($\sim$10$^{36}$~erg~s$^{-1}$);

\item Be/XRTs display a 
high DR$_{1-10~keV}$ ($>$100), 
intermediate duty cycles (DC$_{18-50~keV}$$\sim$10 per cent),
high average  18--50 keV luminosity in outburst ($\sim$10$^{37}$~erg~s$^{-1}$). 
 
\end{itemize}

It is worth mentioning that a number of HMXBs exist that 
displays intermediate properties, in particular among SgHMXB, sometimes overlapping with some 
region of the parameter space more typical of SFXTs, bridging together the two sub-classes.
This seems to indicate that these two sub-classes have no sharp boundaries, 
but their phenomenology is based on continuous parameters, from persistent SgHMXBs towards 
the most extreme SFXT (IGR~J17544-2619).

\begin{discussion}

\discuss{Chaty}{The distinction made with the Corbet diagram between different HMXBs (Be systems,
SgHMXBs and SFXTs) seems not so valid anymore looking at all your plots (SFXTs for instance covering all parameter range).}

\discuss{Sidoli}{Thanks for your comment. Indeed, SFXTs sometimes cover regions typical of both 
Be/XRBs and SgHMXBs.}

\end{discussion}

\end{document}